\documentclass[reprint,amsmath,amssymb,aps,superscriptaddress]{revtex4-1}
\usepackage{amsmath}
\usepackage{hyperref}
\usepackage{graphicx}
\usepackage{bm}
\usepackage[usenames, dvipsnames]{xcolor}

\begin{document}

\title{Neutron elastic scattering on calcium isotopes from chiral nuclear optical potentials}
\author{T. R. Whitehead}
\affiliation{Cyclotron Institute and Department of Physics and Astronomy, Texas A\&M University, College Station, Texas 77843, USA}
\author{Y. Lim}
\address{Max-Planck-Institut f\"ur Kernphysik, Saupfercheckweg 1, 69117 Heidelberg, Germany\\
Institut f\"ur Kernphysik, Technische Universit\"at Darmstadt, 64289 Darmstadt, Germany \\
ExtreMe Matter Institute EMMI, GSI Helmholtzzentrum f\"ur Schwerionenforschung GmbH, 64291 Darmstadt, Germany\\
ylim@theorie.ikp.physik.tu-darmstadt.de}
\author{J. W. Holt}
\affiliation{Cyclotron Institute and Department of Physics and Astronomy, Texas A\&M University, College Station, Texas 77843, USA}

\begin{abstract}
We formulate microscopic neutron-nucleus optical potentials from many-body perturbation theory based on chiral two- and three-body forces. The neutron self-energy is first calculated in homogeneous matter to second order in perturbation theory, which gives the central real and imaginary terms of the optical potential. The real spin-orbit term is calculated separately from the density matrix expansion using the same chiral interaction as in the self-energy. 
Finally, the full neutron-nucleus optical potential is derived within the improved local density approximation utilizing mean field models consistent with the chiral nuclear force employed. We compare the results of the microscopic calculations to phenomenological models and experimental data up to projectile energies of $E = 200$\, MeV. Experimental elastic differential scattering cross sections and vector analyzing powers are generally well reproduced by the chiral optical potential, but we find that total cross sections are overestimated at high energies.
\end{abstract}

\maketitle
\section{Introduction}
Nucleon-nucleus optical potentials describe the interaction of a projectile nucleon with a target nucleus by reducing the complicated many-body interactions to an average single particle potential that is complex and energy-dependent. Global phenomenological optical potentials \cite{Varner91,KD03} are able to describe scattering processes for a large range of nuclei and projectile energies. These global potentials are developed by optimizing their parameters to best reproduce experimental data. Phenomenological potentials yield remarkably good results when interpolating within these ranges, but may not reliably extrapolate to regions where there are no experimental data. Since microscopic optical potentials are built up from fundamental nuclear interactions without tuning to data, they may have greater predictive power in regions of the nuclear chart that are unexplored experimentally.

There has been much interest recently in the development of microscopic optical potentials \cite{Holt13omp,Holt15omp,Egashira,Rotureau18,Rotureau17,Vorabbi18,Toyokawa15,DURANT18,TheBestPaperEver} based on chiral effective field theory (EFT) \cite{WEINBERG79,epelbaum09,MACHLEIDT11}, which implements realistic microphysics including multi-pion exchange processes and three-body interactions all within a framework that allows for the assessment of theoretical uncertainties. Optical potentials based on chiral forces are well suited to describe low-energy nuclear reactions but are expected to break down for energies approaching the cutoff scale of the theory. In practice, the presence of the cutoff constrains nucleon projectile energies to $E \lesssim 200$\,MeV.

In the present study, we compute neutron-nucleus optical potentials along the lines of our previous work in \cite{TheBestPaperEver} that focused exclusively on the description of proton elastic and total reaction cross sections. Since proton elastic scattering at forward angles approaches the well-known Rutherford cross section, the microscopic description of neutron scattering presents a novel challenge that has not yet been addressed in our work. Ultimately our goal is to develop a new microscopic global optical potential for nucleon-nucleus scattering across a large range of isotopes, including those off stability, up to projectile energies of 200 MeV in support of current and future experiments at radioactive ion beam facilities. Presently we consider differential elastic and total cross sections for n-$^{40,48}$Ca scattering at energies ranging from 3-200 MeV. Additionally, in the first direct test of our microscopic spin-orbit term, the vector analyzing power is calculated at selected energies for n-$^{40}$Ca scattering. The choice of isotopes and energies is limited by the availability of experimental data for comparison. We also compare the microscopically calculated scattering observables to the results of the global phenomenological optical potential of Koning and Delaroche \cite{KD03}. Scattering observables are calculated using the TALYS \cite{TALYS} reaction code. While the vector analyzing power by is not output directly by TALYS, it can be extracted from the output files of ECIS-06, a program that runs in the background of TALYS.

We take as the foundation of our calculations a high-precision 2N + 3N chiral nuclear interaction with a momentum-space cutoff of $\Lambda = 450$\,MeV. The low-energy constants of the potential are fitted to empirical data. For two-body interactions, the empirical inputs include nucleon-nucleon scattering phase shifts and deuteron properties. Three-body contact terms are fit to the triton binding energy and lifetime \cite{coraggio14}. The nucleon-nucleon interaction is calculated to next-to-next-to-next-to-leading order (N3LO), while the three-nucleon force is only calculated at N2LO. Work towards the inclusion of three-nucleon N3LO interactions is in progress \cite{tews13,drischler16,kaiser18,kaiser19,drischler19,holt20} and we plan to implement them in future works. The chiral nuclear potential employed in the present work reproduces known values for nuclear matter properties, such as saturation energy and density \cite{coraggio14}, thermodynamics \cite{wellenhofer14,wellenhofer15}, and Fermi liquid parameters \cite{Holt18} when calculated to at least second order in many-body perturbation theory. In future works we also plan to calculate the nucleon-nucleus optical potential with a selection of high-precision chiral nuclear forces \cite{entem17,reinert18} to better assess theoretical uncertainties.

In quantum many-body theory, the energy- and momentum-dependent single-particle self-energy is equivalent to the optical potential for scattered states \cite{PhysRevLett.3.96}. We first compute the nucleon self-energy in homogeneous nuclear matter of arbitrary density and proton fraction to second order in many-body perturbation theory including chiral two- and three-body forces. We next compute nuclear density distributions for $^{40,48}$Ca from mean field theory employing the Sk${\chi}$450 Skyrme effective interaction \cite{Lim17} constrained by chiral EFT. In the local density approximation (LDA) the nucleon-nucleus optical potential is computed \cite{Jeukenne77lda} by folding the nuclear matter optical potential with a nuclear density distribution. The LDA is known to underestimate the surface diffuseness of the optical potential in finite nuclei and requires a modification known as the improved local density approximation (ILDA) \cite{Jeukenne77lda,DelarocheILDA} that accounts for the nonzero range of the nuclear interaction. 

The main advantage of the nuclear matter approach to deriving optical potentials is its adaptability to many nuclei. Once the nuclear matter optical potential is calculated, only the nuclear density distribution is needed to produce a nucleon-nucleus optical potential, making the nuclear matter approach well suited to constructing a {\it microscopic global} optical potential. However, the drawback is that some physical processes present in scattering with finite nuclei are not captured by nuclear matter calculations. Among these are collective surface modes, shell structure effects, and the fact that the spin-orbit term is not present in homogeneous nuclear matter. We therefore include a microscopic spin-orbit term from the improved density matrix expansion \cite{bogner08,gebremariam10,KaiserHoltEDF} based on chiral interactions. Compared to the standard density matrix expansion of Negele and Vautherin \cite{negele72}, the improved density matrix expansion provides a better description of the spin-dependent part of the energy density functional.

\begin{figure}[t]
\includegraphics[scale=0.45]{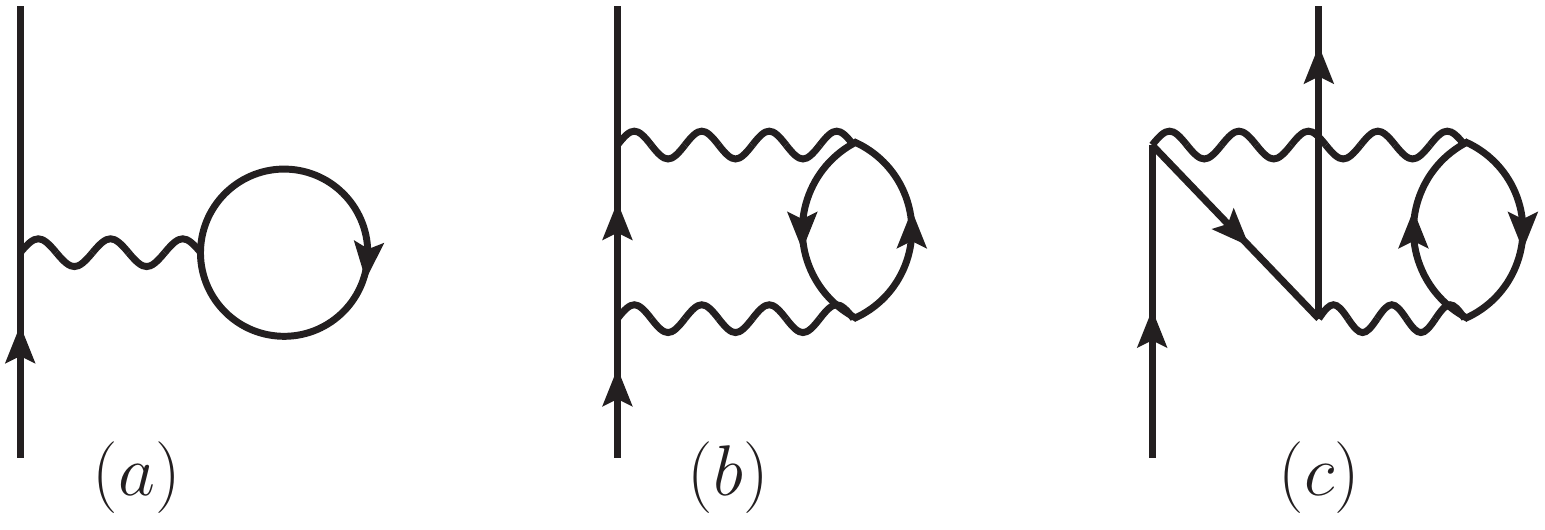}
\caption{The first and second order contributions to the self-energy represented diagrammatically. Nucleon propagators are represented by solid lines and the the in-medium two-nucleon interaction is represented by wavy lines.}
\label{fig:diagrams}
\end{figure}

The paper is organized as follows. In Section \ref{chiral} we calculate the microscopic optical potential in nuclear matter with nucleon interactions from chiral EFT. We then calculate nuclear density distributions from mean field theory with a Skyrme interaction fit to the chiral EFT potential used in the self-energy. The ILDA is then employed to construct nucleon-nucleus optical potentials for $^{40,48}$Ca. Finally, the microscopic optical potentials are parameterized to the Koning-Delaroche (KD) phenomenological form in order to implement them in the reaction code TALYS. In Section \ref{results} we compute neutron-nucleus elastic differential scattering cross sections up to a projectile energy $E = 185$\,MeV and total cross sections up to $E=200$\,MeV. We also calculate the vector analyzing power for elastic n-$^{40}$Ca scattering as a test of our spin-orbit term. These results are compared to experiment and predictions of the KD phenomenological model. We end with a summary and conclusions.


\section{Optical potential from chiral effective field theory}
\label{chiral}

\subsection{Real and imaginary central terms}
The nucleon self-energy is calculated as a function of density and momentum in homogeneous nuclear matter of arbitrary isospin asymmetry using a nuclear potential derived from chiral EFT. The expressions for the first- and second-order perturbative contributions to the nucleon self-energy are given by
\begin{equation}
\Sigma^{(1)}_{2N}(q;k_f)=\sum_{1} \langle \vec{q} \: \vec{h}_1 s s_1 t t_1 | \bar{V}_{2N}^{\rm eff} | \vec{q} \: \vec{h}_1 s s_1 t t_1 \rangle n_1 ,
\label{eq:2}
\end{equation}
\begin{eqnarray}
\label{sig2a}
&&\hspace{-.1in} \Sigma^{(2a)}_{2N}(q,\omega;k_f) \\ \nonumber
&&= \frac{1}{2} \sum_{123} \frac{|\langle \vec{p}_1 \vec{p}_3 s_1 s_3 t_1 t_3 | \bar{V}_{2N}^{\rm eff} | \vec{q} \vec{h}_2 s s_2 t t_2 \rangle|^2}{\omega+\epsilon_2-\epsilon_1-\epsilon_3+i\eta} \bar{n}_1 n_2 \bar{n}_3,
\end{eqnarray}
\begin{eqnarray}
\label{sig2b}
&&\hspace{-.1in} \Sigma^{(2b)}_{2N}(q,\omega;k_f) \\ \nonumber
&&= \frac{1}{2} \sum_{123} \frac{|\langle \vec{h}_1 \vec{h}_3 s_1 s_3 t_1 t_3 | \bar{V}_{2N}^{\rm eff} | \vec{q} \vec{p}_2 s s_2 t t_2 \rangle|^2}{\omega+\epsilon_2-\epsilon_1-\epsilon_3-i\eta}   n_1 \bar{n}_2 n_3,
\end{eqnarray}
and shown diagrammatically in Fig.\ \ref{fig:diagrams}
In the above expressions $n_i=\theta(k_f - k_i)$ is the occupation probability for a filled state with momentum $ k_i < k_f$ below the Fermi momentum, 
and particle state occupation probabilities are given by $\bar n_i = \theta(k_i-k_f)$, with the summation going over intermediate-state momenta for particles $\vec p_i$ and holes $\vec h_i$, their spins $s_i$, and isospins $t_i$. The overbar on the potential indicates that it is properly antisymmetrized. The in-medium effective nuclear potential $V_{2N}^{\rm eff}$ is the two-body interaction which consists of the bare nucleon-nucleon (NN) potential $V_{NN}$ plus an effective, density-dependent (and isospin-asymmetry-dependent) NN interaction $V_{NN}^{\rm med}$ derived from the N2LO chiral three-nucleon force by averaging one particle over the filled Fermi sea of noninteracting nucleons \cite{bogner05,Holt09,hebeler10,holt20}. In the first-order Hartree-Fock contribution, Eq.\ (\ref{eq:2}), the effective interaction is given by $V_{2N}^{\rm eff} = V_{NN} + \frac{1}{2}V_{NN}^{\rm med}$, while for the higher-order contributions, Eqs.\ (\ref{sig2a}) and (\ref{sig2b}), the effective interaction is given by $V_{2N}^{\rm eff} = V_{NN} + V_{NN}^{\rm med}$. The single-particle energies in the denominators of Eqs.\ (\ref{sig2a}) and (\ref{sig2b}) are computed self-consistently according to
\begin{equation}
\label{eq:1}
\epsilon(q)=\frac{q^2}{2M}+\text{Re} \Sigma(q,\epsilon(q)),
\end{equation}
where $M$ is the free-space nucleon mass.

The Hartree-Fock contribution shown in Fig.\ \ref{fig:diagrams} (a) represents the mean field interaction of a propagating nucleon with each of the constituent particles in the medium. It is nonlocal, energy independent, and real. The second-order contributions shown in Fig.\ \ref{fig:diagrams} (b) and (c) represent the effects of virtual particle-hole excitations of the medium, which in general produce nonlocal, energy dependent, and complex self energies. In particular, Fig.\ \ref{fig:diagrams} (b) gives rise to an imaginary part when $q > k_f$, while Fig.\ \ref{fig:diagrams} (b) gives rise to an imaginary part when $q < k_f$. The relative strength of the different contributions depends to some extent on the resolution scale of the nuclear potential, with soft interactions generically shifting strength from the second-order to first-order contributions. In all cases considered here, the first-order Hartree-Fock contribution from two-body forces is attractive, while that from three-body forces is repulsive. The self-energy from three-body forces grows approximately quadratically with the density \cite{holt20}, stronger than for two-body forces alone. For coarse resolution chiral potentials, the Hartree-Fock contribution is larger than the second-order terms, indicating improved convergence in the perturbation series expansion.

To derive optical potentials for neutron- or proton-rich nuclei, it is necessary to calculate the self-energy for arbitrary isospin-asymmetry, $\delta_{np} = (\rho_n-\rho_p)/(\rho_n+\rho_p)$. Both the real and imaginary terms of the chiral optical potential have significant isovector components for proton and neutron projectiles. Effects of the isovector contribution to the optical potential will be discussed in more detail in later sections. The resulting optical potentials for nucleons propagating in homogeneous matter with proton and neutron Fermi momenta $k_f^p$ and $k_f^n$ are given by

\begin{eqnarray}
U_p(E;k_f^p,k_f^n) &=& V_p(E;k_f^p,k_f^n) + i W_p(E;k_f^p,k_f^n), \nonumber \\ 
U_n(E;k_f^p,k_f^n) &=& V_n(E;k_f^p,k_f^n) + i W_n(E;k_f^p,k_f^n)
\label{omp}
\end{eqnarray}
with
\begin{equation}
V_i(E;k_f^p,k_f^n) = {\rm Re}\Sigma_i(q,E(q);k_f^p,k_f^n),
\end{equation}
\begin{equation}
W_i(E;k_f^p,k_f^n) = \frac{M_i^{k*}}{M} {\rm Im}\Sigma_i(q,E(q);k_f^p,k_f^n),
\end{equation}
where the subscript $i$ denotes a propagating proton or neutron. To relate the microscopically derived imaginary part of the nucleon self-energy to the imaginary term of the optical potential used in phenomenology, the non-locality must be accounted for \cite{Negele81,fantoni81}. This is achieved by multiplying the imaginary term of the self-energy by the effective $k$-mass $M_i^{k*}$ defined by
\begin{equation}
\frac{M_i^{k*}}{M} = \left ( 1 + \frac{M}{k}\frac{\partial}{\partial k}V_i(k,E(k) \right )^{-1}.
\end{equation}

\subsection{Spin-orbit optical potential}
The effective one-body spin-orbit interaction vanishes in homogeneous nuclear matter due to translational invariance and thus cannot be computed within the framework described above. Alternatively, we employ an improved density matrix expansion \cite{gebremariam10,Gebremariam10npa,KaiserHoltEDF} to construct the one-body spin-orbit interaction from chiral two- and three-body forces. By utilizing the improved density matrix expansion that takes advantage of phase space averaging, a more accurate spin-dependent energy density functional can be derived compared to the standard density matrix expansion of Negele-Vautherin \cite{negele72}. 

The density matrix is defined by
\begin{equation}
\rho(\vec r_1 \sigma_1 \tau_1; \vec r_2 \sigma_2 \tau_2) = \sum_\alpha \Psi_\alpha^*(\vec r_2 \sigma_2 \tau_2)
\Psi_\alpha(\vec r_1 \sigma_1 \tau_1),
\end{equation}
where $\Psi_\alpha$ are the energy eigenfunctions of the occupied orbitals in the non-relativistic many-body system. The energy density functional for a $N=Z$ even-even nucleus in the Hartree-Fock approximation expanded to second order in spatial gradients is given by
\begin{eqnarray}
{\cal E}[\rho,\tau,\vec J\,] &=& \rho\,\bar E(\rho)+\bigg[\tau-
{3\over 5} \rho k_f^2\bigg] \bigg[{1\over 2M_N}+F_\tau(\rho)
\bigg] \\  \nonumber && + (\vec \nabla \rho)^2\, F_\nabla(\rho)+  \vec \nabla
\rho \cdot\vec J\, F_{SO}(\rho)+ \vec J\,^2 \, F_J(\rho)\, ,
\label{edf}
\end{eqnarray}
where $\rho(\vec r\,) =2k_f^3(\vec r\,)/3\pi^2 =  \sum_\alpha \Psi^\dagger_\alpha(\vec r\,) \Psi_\alpha(\vec r\,)$ is the local density and $k_f(\vec r\,)$ is the local Fermi momentum. The kinetic energy density is given by $\tau(\vec r\,) =  \sum_\alpha \vec \nabla \Psi^\dagger_\alpha (\vec r\,) \cdot \vec \nabla \Psi_\alpha(\vec r\,)$ and the spin-orbit density is given by $\vec J(\vec r\,) = i \sum_\alpha \vec \Psi^\dagger_\alpha(\vec r\,) \vec \sigma \times \vec \nabla \Psi_\alpha(\vec r\,)$. This calculation yields the spin-orbit term $F_{SO}(\rho)$ of the optical potential for $N=Z$ nuclei to first order in many-body perturbation theory. Higher-order perturbative contributions \cite{zhang18} to the microscopic nuclear energy density functional will be investigated in future works. The isovector part \cite{Kaiser12} of the spin-orbit interaction for $^{48}$Ca is not included in this study since it is small compared to the isoscalar part \cite{KaiserHoltEDF}. In the context of nucleon-nucleus scattering, the spin-orbit term of the optical potential determines the polarization of scattered nucleons. One such polarization observable is the vector analyzing power, which we will calculate microscopically and compare to experimental data and phenomenological results.

\subsection{Improved local density approximation}

\begin{figure}[t]
\includegraphics[scale=0.47]{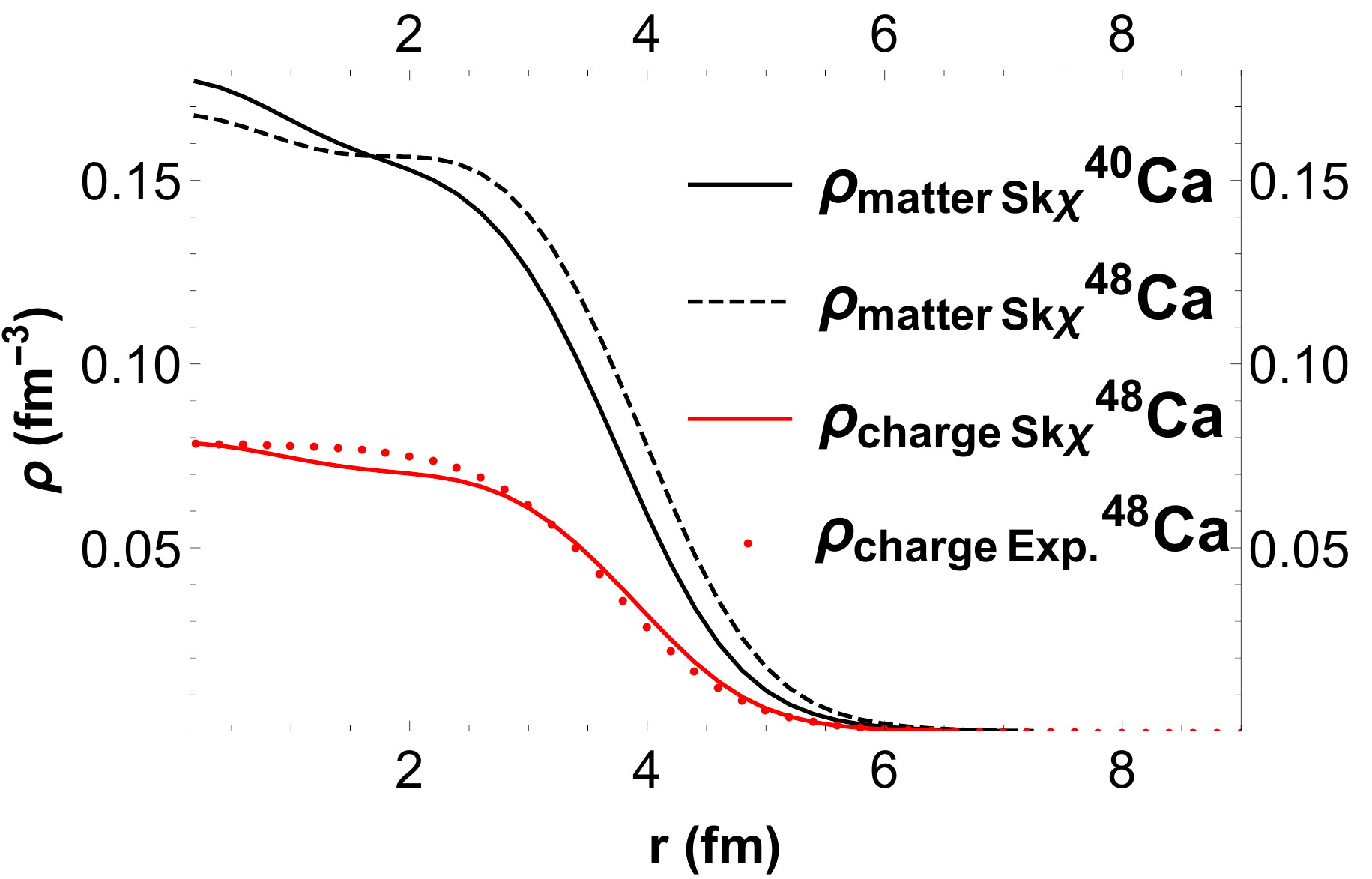}
\caption{The matter density distributions for $^{40,48}$Ca, represented by a black curve and a dashed black curve respectively, are calculated in mean field theory from the Skyrme Sk$\chi$450 effective interaction constrained by chiral effective field theory. The empirical charge density distribution for $^{48}$Ca along with the mean field calculation are represented by red dots and a red curve respectively.}
\label{fig:Ca40denprof}
\end{figure}

The improved local density approximation (ILDA) is used to construct the nucleon-nucleus optical potential from the nucleon self-energy in nuclear matter. The nucleon-nucleus optical potential is derived by folding the density-dependent self-energy with the radial density distribution of a target nucleus and then smeared by integrating over the radial dimension with a Gaussian factor to account for the nonzero range of the nuclear force. We calculate the nuclear density distributions within mean field theory from the Sk$\chi$450 Skyrme interaction \cite{Lim17}. The Sk$\chi$450 interaction is fit to properties of finite nuclei in addition to theoretical calculations of the asymmetric nuclear matter equation of state from the N3LO chiral potential with cutoff scale $\Lambda = 450$\,MeV used to calculate the self-energy. In Fig.\ \ref{fig:Ca40denprof} we show the calculated nucleon density distributions for $^{40,48}$Ca. In order to benchmark these density distributions with experiment we also show the charge density distribution for $^{48}$Ca calculated from mean field theory compared to an empirical charge density \cite{DEVRIES1987495} obtained from electron scattering data. The theoretical charge density for $^{48}$Ca slightly underestimates experimental results from  $1\,{\rm fm} < r < 3 \,{\rm fm}$ and slightly overestimates in the surface region. We have verified as well that the charge density of $^{40}$Ca from mean field theory has a qualitatively similar comparison to experiment.

In the local density approximation, the nucleon-nucleus optical potential is evaluated as
\begin{eqnarray}
&&\hspace{-.2in}V(E;r) + i W(E;r) = V(E;k_f^p(r),k_f^n(r)) \\ \nonumber
&& + i W(E;k_f^p(r),k_f^n(r)),
\end{eqnarray}
where $k_f^p(r)$ and $k_f^n(r)$ are the local proton and neutron Fermi momenta.
This approximation does not account for the nonzero range of nuclear forces, and is known to underestimate the surface diffuseness of nucleon-nucleus optical potentials \cite{BRIEVA1977317,Jeukenne77lda}. For this reason, the standard LDA provides an inadequate description of nuclear scattering processes. To account for the range of the nuclear force and obtain a more realistic nuclear optical potential, the improved local density approximation is employed. The ILDA applies a Gaussian smearing
\begin{equation}\label{eq:ilda}
{V}(E;r)_{ILDA}=\frac{1}{(t\sqrt{\pi})^3}\int V(E;r') e^{\frac{-|\vec{r}-\vec{r}'|^2}{t^2}} d^3r'
\end{equation}
that is characterized by a variable length scale $t$ associated with the range of the nuclear force. In Ref.\ \cite{DelarocheILDA} it is found that for the central part of the optical potential $t_C=1.3\,{\rm fm}$ gives the best fit to experimental neutron total cross sections for $10\,{\rm MeV} < E < 200 \,{\rm MeV}$ and targets ranging from $^{40}$Ca to $^{208}$Pb. Presently we vary the range parameter over $1.25\,{\rm fm} < t_C < 1.35\,{\rm fm}$ to estimate the theoretical uncertainty associated with the choice of length scale $t_C$. As in \cite{TheBestPaperEver}, we find the spin-orbit range parameter to be $t_{SO}=1.07$ fm and vary it across the range $1.0\,{\rm fm} < t_{SO} < 1.1\,{\rm fm}$ to estimate the uncertainty. 

Several alternative approaches, such as Watson multiple scattering theory \cite{vorabbi16} 
and the $G$-matrix folding method \cite{Toyokawa15}, also employ nuclear density 
distributions folded with the nuclear interaction to produce microscopic optical potentials. 
Currently, Pauli-blocking effects and three-body forces are challenging to implement in a 
full multiple scattering formalism \cite{vorabbi16}, but the $G$-matrix folding method has the flexibility to
include these effects. In the $G$-matrix folding approach, the local density is evaluated at the 
midpoint of the incident projectile and the target density element, integrated over the entire target
volume to generate the optical potential. Compared to the ILDA employed in the present work,
the advantage of the folding prescription in the $G$-matrix approach is that no adjustable 
smearing factor needs to be introduced, but the disadvantage is that the full finite-range 
character of the nuclear force may be underestimated. Qualitatively, the two methods give 
similar effects when three-body forces are introduced, 
namely a reduction in the strength of the real part and 
an enhancement of the absorptive imaginary part at high energies \cite{Toyokawa15} as 
we now show. 

\begin{figure*}[t]
\includegraphics[scale=0.32]{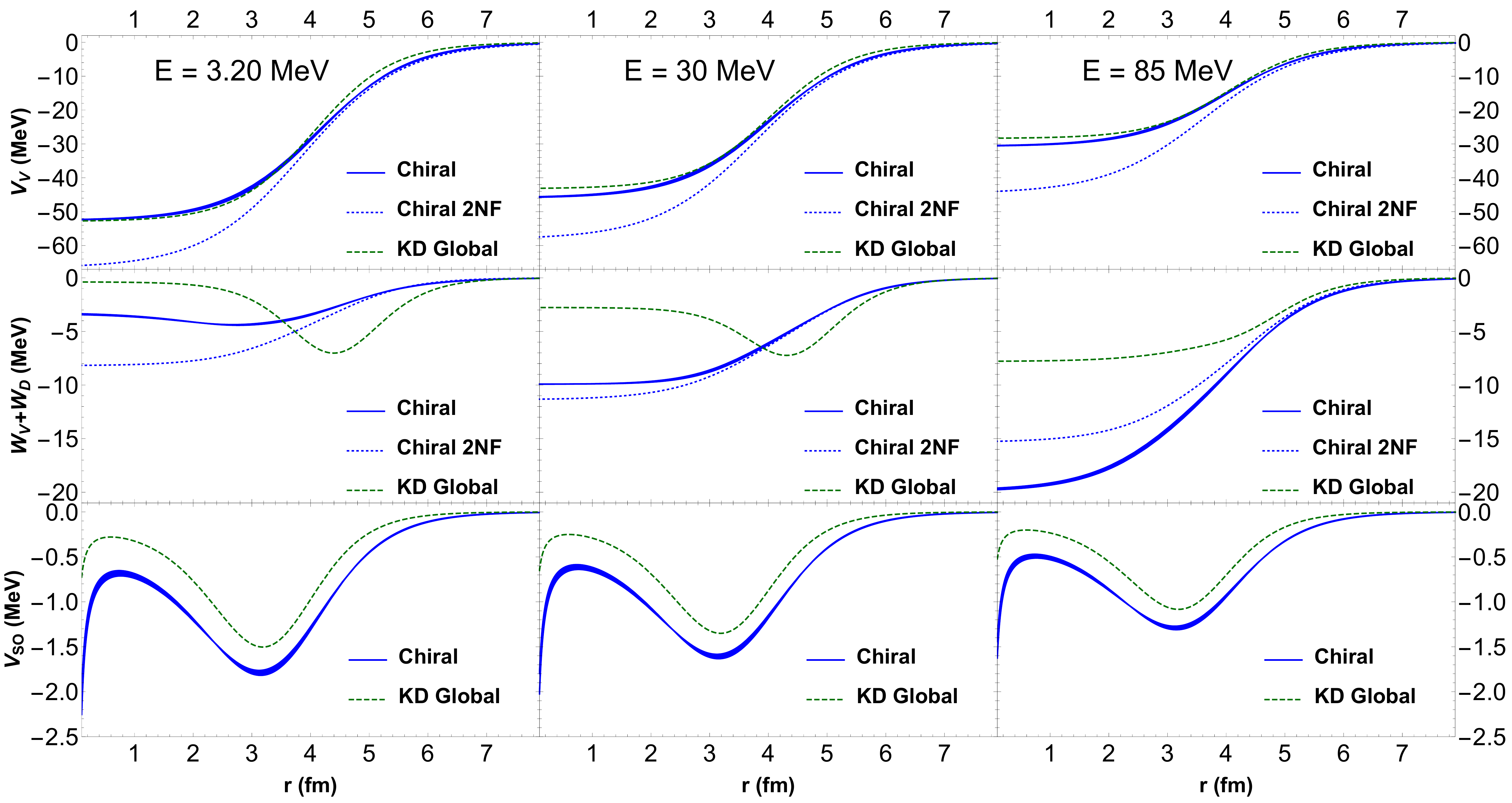}
\caption{The real, imaginary, and spin-orbit terms of the n-$^{40}$Ca optical potential at projectile energies $E=3.20, 30, 85$\,MeV. The blue bands represent the microscopic chiral optical potential after applying the improved local density approximation with a varied length scale, and the blue dotted line represents the chiral potential with only two body forces. The green dashed lines represent the analogous terms of the Koning-Delaroche global optical potential.}
\label{fig:potentialplots}
\end{figure*}

In Fig.\ \ref{fig:potentialplots} we show the real central, imaginary central, and real spin-orbit terms of the ILDA chiral optical potential compared to the analogous terms of the KD phenomenological optical potential for n-$^{40}$Ca at projectile energies $E=3.2, 30, 85$\,MeV. The width of the blue band representing the chiral terms shows the relatively small effect of varying the distance parameter in the ILDA. In the left column of plots, the optical potential terms are shown at $E=3.2$\,MeV. The microscopic real volume term has a very similar depth and a slightly larger diffuseness compared to the KD term. At this low energy, the microscopic imaginary term has a surface peak and a nonzero central depth, whereas the KD imaginary term has virtually no central depth and a relatively large surface peak. The microscopic spin-orbit term has a very similar radial profile compared to KD, but with a larger depth across all energies. The density matrix expansion calculated at the Hartree-Fock level is known \cite{KaiserHoltEDF,holt16pr} to produce a stronger spin-orbit interaction than is required from traditional mean field theory studies of finite nuclei by about 20-50\%. The inclusion of multi-pion-exchange processes has been shown \cite{kaiser10} to reduce the strength of the one-body spin-orbit interaction in finite nuclei. In future works we intend to account for these processes by including $G$-matrix correlations in the density matrix expansion as outlined in Ref.\ \cite{zhang18}. 

At $E=30$\,MeV the middle column of plots in Fig.\ \ref{fig:potentialplots} shows a microscopic real volume term that has a slightly larger central depth and similar diffuseness compared to phenomenology. The microscopic imaginary term has a large central depth with almost no surface peak, while its phenomenological counterpart has a small central depth and moderate surface peak. This feature has been observed in other microscopic optical potentials calculated from nuclear matter \cite{LEJEUNE78,Lagrange82,Kohno84,Petler85}. To mitigate this discrepancy, some semi-microscopic optical potentials apply an energy-dependent scaling factor to the imaginary term \cite{DelarocheILDA,goriely07}, but in the present work we employ no such factors. As the energy increases to $E=85$\,MeV, the real volume term becomes more shallow for both the microscopic and phenomenological potentials while qualitatively remaining the same relative to each other. At such high energy, the imaginary surface peak is no longer present in either the microscopic or phenomenological potentials. However, at this energy the central depth of the microscopic imaginary term is very large compared to phenomenology. This results in a chiral optical potential that is overly absorptive at high energy. 

In Fig.\ \ref{fig:potentialplots} we also show the real and imaginary terms of the chiral optical potential with only two-body forces included. The contribution of three-body forces is most evident in the central region where densities are largest. As the density decreases towards the surface, three-body contributions become smaller and the two chiral potentials converge. For all three energies, three-body forces reduce the central depth of the real part by $\sim$ 12 \,MeV. This reduction of the depth by the three-body interactions is due to their generally repulsive nature. However, for the imaginary term, three-body forces reduce the depth only for energies $E<45$\,MeV. Beyond this energy, three-body contributions increase the depth by a growing amount as energy increases. This has been observed in previous work \cite{Toyokawa15}, and we find good qualitative agreement for the chiral imaginary terms at $E=85$\,MeV compare Ref.\ \cite{Toyokawa15}. The large contribution of three-body forces at N2LO in the chiral expansion further motivates the future inclusion of N3LO three-body forces in our calculations \cite{holt20}.

\subsection{Parameterization of the chiral optical potential}
We fit our optical potentials to the phenomenological form of Koning and Delaroche in order to implement them in nuclear reaction codes. We aim to eventually construct a global microscopic optical potential based on chiral forces and make it available in a convenient form for the nuclear reaction community. The Koning-Delaroche phenomenological neutron optical potential takes the form
\begin{eqnarray}
&&\hspace{-.3in} U(r,E) = V_V(r,E) + i W_V(r,E) + i W_D(r,E) \label{phen} \\ \nonumber 
&& + V_{SO}(r,E) \vec \ell \cdot \vec s + i W_{SO}(r,E) \vec \ell \cdot \vec s,
\end{eqnarray} 
consisting of a real volume term, imaginary volume term, imaginary surface term, and real and imaginary spin-orbit terms. The imaginary spin-orbit term is not considered in the current work since it has a negligible effect on elastic scattering cross sections at low energies \cite{BRIEVA1978206} due to its very small magnitude. Furthermore, it cannot be extracted within the present microscopic approach. The terms of the phenomenological optical potential are assumed to have energy and radial dependences that factorize according to
\begin{equation}
V_V(r,E) = {\cal V}_V(E) f(r;r_V,a_V),
\label{vreal}
\end{equation}
\begin{equation}
W_V(r,E) = {\cal W}_V(E) f(r;r_W,a_W),
\label{vim}
\end{equation}
\begin{equation}
W_D(r,E) = -4a_D{\cal W}_D(E) \frac{d}{dr} f(r;r_D,a_D),
\label{sim}
\end{equation}
\begin{equation}
V_{SO}(r,E) = {\cal V}_{SO}(E) \frac{1}{m_\pi^2}\frac{1}{r}\frac{d}{dr} f(r;r_{SO},a_{SO}),
\label{soreal}
\end{equation}
where
\begin{equation}
f(r;r_i,a_i) = \frac{1}{1+e^{(r-A^{1/3}r_i)/a_i}}
\end{equation}
is the Woods-Saxon shape factor with $A$ the mass number and $r_i$,$a_i$ the energy-independent geometry parameters that represent the size and diffuseness of a given target nucleus respectively. In phenomenological and microscopic optical potentials, these shape parameters vary weakly as functions of the neutron $N$ and proton $Z$ numbers. The chiral optical potential is fit to the KD form at a given energy thus there is no explicit parameterization of the energy dependence. We note that the microscopic real spin-orbit optical potential is calculated from the density matrix expansion at the Fermi energy $E_F$ and has no energy dependence. We therefore incorporate a phenomenological energy dependence that is small and constant across all nuclei into our parameterization of the spin-orbit optical potential.

\section{Results}
\label{results}

\begin{figure}[t]
\includegraphics[scale=0.35]{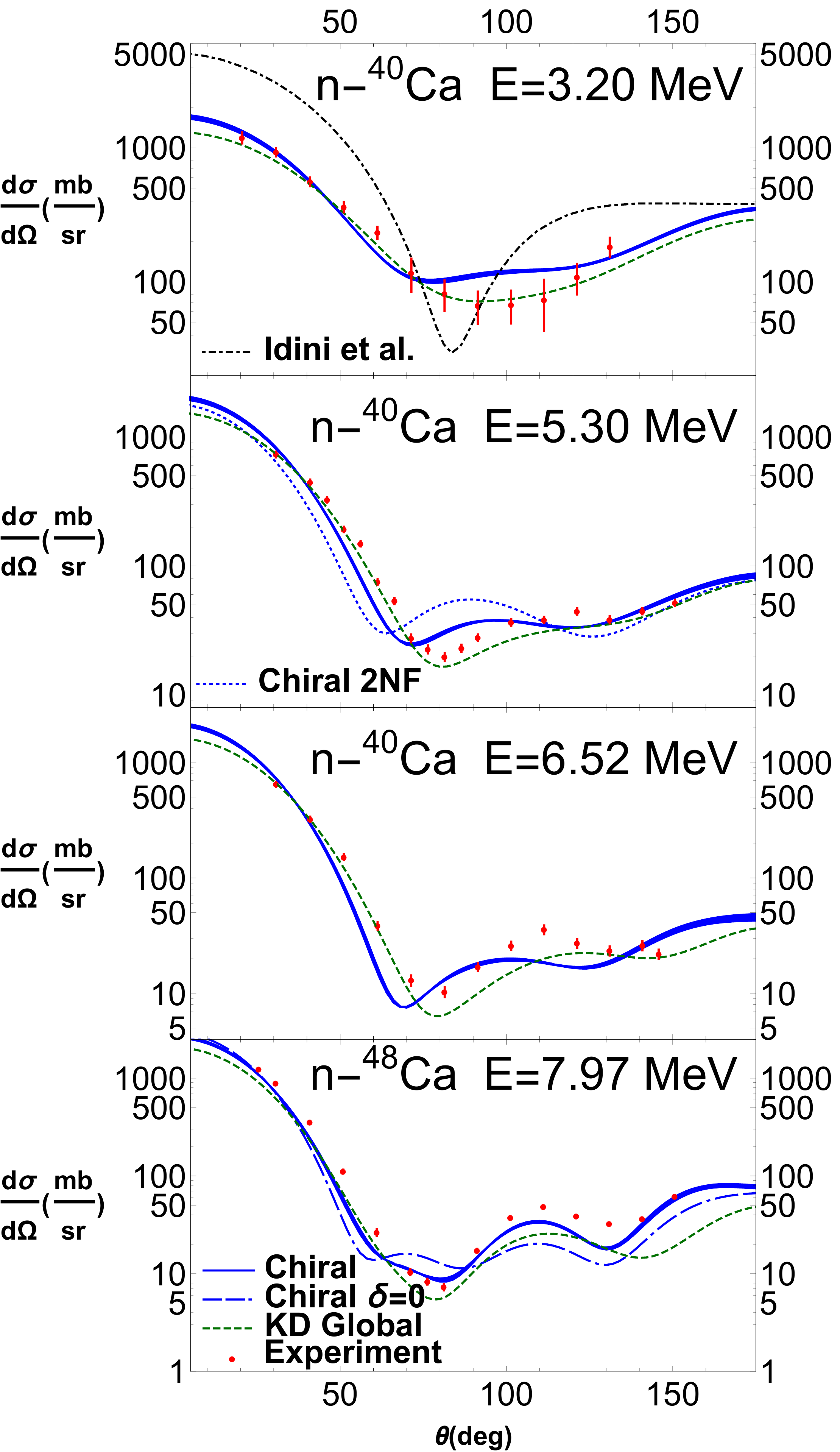}
\caption{Differential elastic scattering cross sections for n-$^{40}$Ca at projectile energies $E=3.2. 5.3, 6.52$\,MeV and n-$^{48}$Ca at $E=7.97$\,MeV. Cross sections calculated from the chiral optical potential are given by blue bands. Cross sections calculated from the Koning-Delaroche phenomenological optical potential are given by green dashed curves, and experimental data are represented by red circles with error bars. In the top plot, the dot dashed black curve represents ab initio calculations found in Ref.\ \cite{PhysRevLett.123.092501}. In the plot second from the top, the blue dotted curve is calculated from the chiral optical potential with only two nucleon forces (2NF). In the bottom plot, the blue dash-dash-dot curve is calculated from symmetric nuclear matter ($\delta=0$) to demonstrate the effect of the isovector component of the optical potential.}
\label{csplot1}
\end{figure}

\begin{figure*}[t]
\includegraphics[scale=0.35]{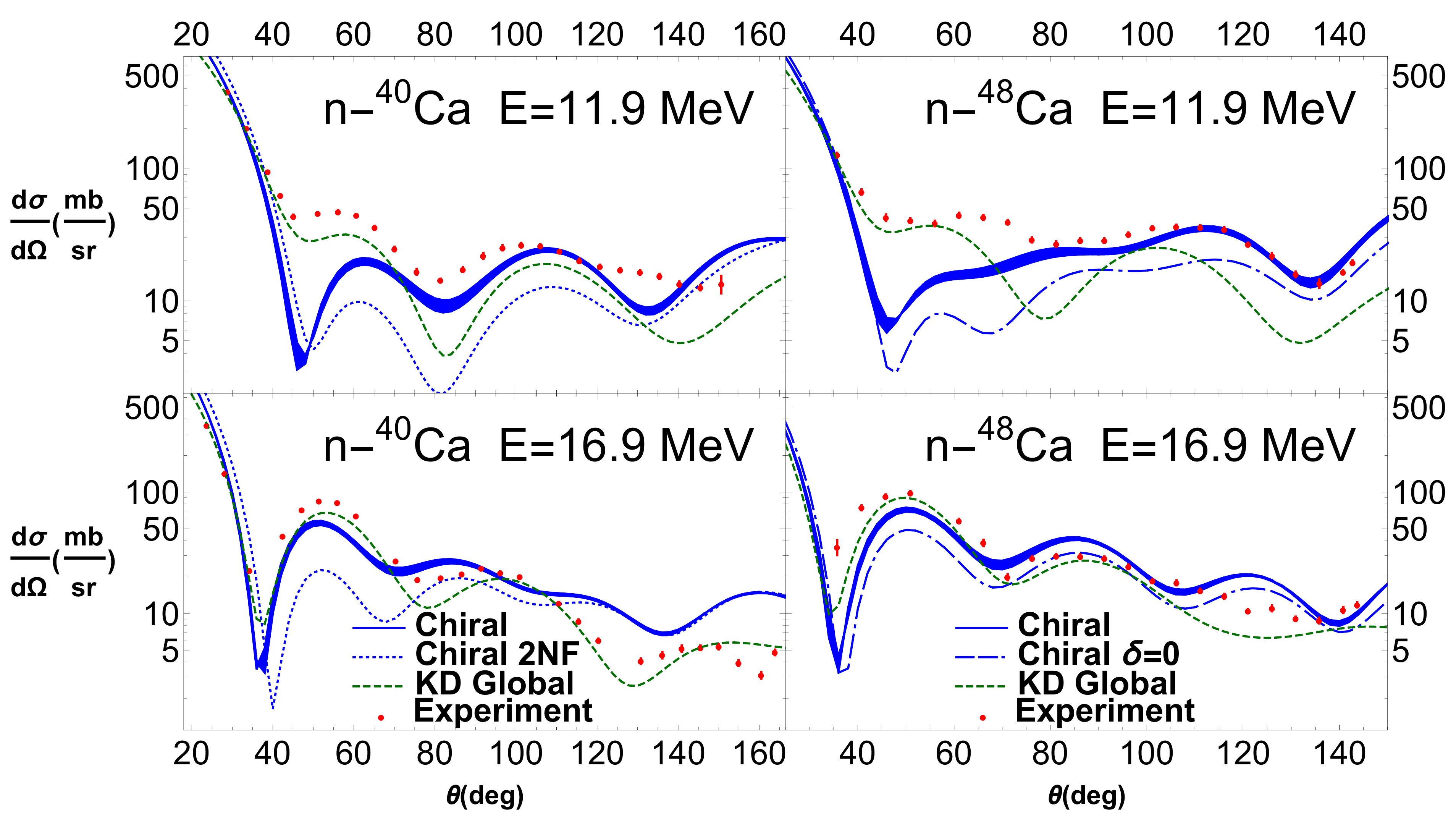}
\caption{Differential elastic scattering cross sections for n-$^{40,48}$Ca at projectile energies $E=11.9, 16.9$\,MeV. Cross sections calculated from the chiral optical potential are given by blue bands. Cross sections calculated from the Koning-Delaroche phenomenological optical potential are given by green dashed curves, and experimental data are represented by red circles with error bars. In the left two plots, the blue dotted curves are calculated from the chiral optical potential with only two nucleon forces (2NF), and in the right plots the blue dash-dash-dot curves are calculated from symmetric nuclear matter ($\delta=0$) to demonstrate the effect of the isovector component of the optical potential.}
\label{csplot2}
\end{figure*}

\begin{figure}[t]
\includegraphics[scale=0.37]{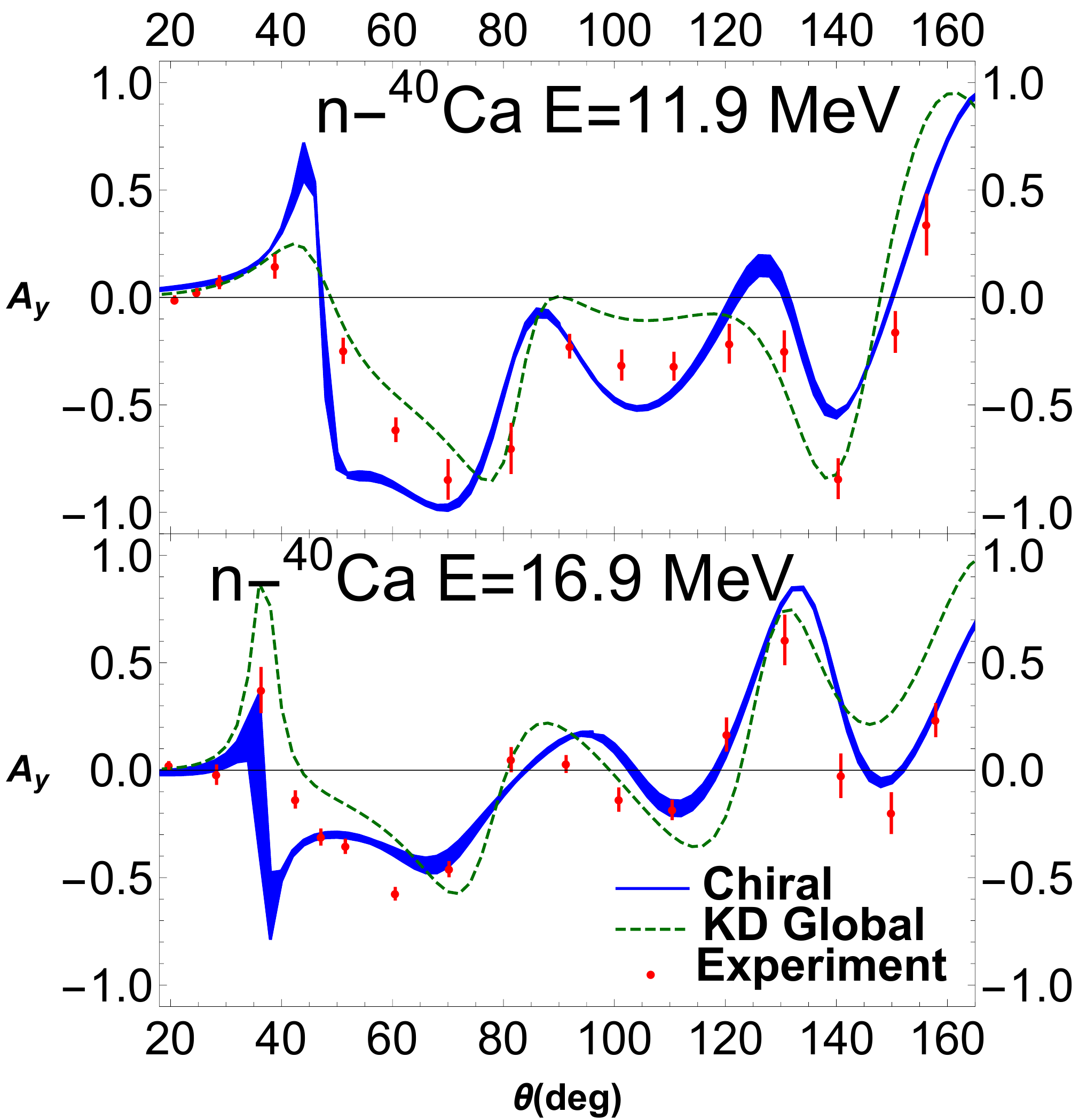}
\caption{Vector analyzing powers for elastic n-$^{40}$Ca scattering at projectile energies $E=11.9, 16.9$\,MeV. Vector analyzing powers calculated from the chiral optical potential are given by the blue bands. Vector analyzing powers calculated from the Koning-Delaroche phenomenological optical potential are given by the green dashed curves, and experimental data are represented by red circles with error bars.}
\label{ayplot}
\end{figure}

In a continuation of Ref.\ \cite{TheBestPaperEver}, we calculate cross sections and vector analyzing powers of neutrons scattering on calcium isotopes from a microscopic optical potential based on chiral forces and compare to experiment and phenomenology. We also include elastic cross sections calculated from the chiral optical potential with only two-body interactions, and in the case of $^{48}$Ca, without the isovector term to show the importance of including three-body forces and isospin asymmetry in the calculation of the optical potential.

Both the differential elastic scattering cross sections and total cross sections are calculated for n-$^{40,48}$Ca at energies where experimental data are available. Specifically, we compute differential elastic scattering cross sections for n+$^{40}$Ca at projectile energies $E=$ 3.2, 5.3, 6.52, 11.9, 16.9, 21.7, 25.5, 30, 40, 65, 85, 107.5, 155, 185\,MeV. In order to test the spin-orbit term, vector analyzing powers are also calculated at $E=11.9, 16.9$\,MeV for n-$^{40}$Ca. Differential elastic scattering cross sections are calculated for n-$^{48}$Ca at $E=7.97, 11.9, 16.9$\,MeV. The total cross sections for n-$^{40,48}$Ca scattering are also calculated. Energies greater than 200 \,MeV are not considered since the chiral expansion is expected to break down near that energy scale \cite{holt20}. Experimental data are taken from Refs.\ \cite{Ca48e7,Ca48e12and17,Ca40e65,Ca40e21and25,Ca40e30and40,Ca40e5and6,Ca40eHigh,Ca40e17,Ca40e12,Ca40e3,TotalCS}.

The TALYS reaction code is used to calculate all scattering observables. In all cases we employ the microscopic optical potential calculated using the ILDA and parameterized to the Koning-Delaroche phenomenological form at a specific energy. Presently the only theoretical uncertainties considered are the variations in the ILDA length scales $t_C$ and $t_{SO}$. In future works we will consider multiple chiral nuclear potentials of varying order and cutoff in order to more accurately assess the complete theoretical uncertainty. We also include results from the KD global phenomenological optical potential \cite{KD03}. 

Compared to the proton-calcium elastic scattering cross sections found in Ref.\ \cite{TheBestPaperEver}, we find that neutron-calcium cross sections are in somewhat better agreement with experiment, especially at backward scattering angles. This difference is possibly due to the approximate treatment of the Coulomb interaction in the proton-nucleus optical potential. In TALYS the charge distribution of the target nucleus is assumed to be that of a uniformly-charged sphere. This approximation may provide an inadequate description for high momentum-transfer interactions in which backscattered nucleons probe the interior of the nucleus. For both neutrons and protons, we find that the chiral optical potential reproduces elastic cross sections very well at low energies, $E \leq 10$\,MeV, and provides an adequate description up to energies of $E \sim 200$\,MeV.

\subsection{Microscopic optical potential at low energy}
Low-energy nuclear reactions are important for describing a wide range of astrophysical processes. These reactions play an essential role in cold $r$-process environments \cite{mumpower15,horowitz19} such as neutron star mergers where freeze-out is achieved rapidly and neutron capture plays an enhanced role. Neutron capture rates on exotic, neutron-rich isotopes have large theoretical uncertainties \cite{mumpower15}. These neutron-capture rates are included as inputs for most modern $r$-process reaction network codes. The neutron-nucleus optical potential, and especially the imaginary part of the optical potential at low energies \cite{goriely07}, is a key ingredient in calculating neutron capture rates. One of the primary motivations for the construction of a new global microscopic optical potential is to better understand (and potentially reduce) these theoretical uncertainties. In the future, we will directly implement the developed microscopic optical potentials to applications including neutron-capture cross sections. In the present work, we benchmark to differential elastic scattering at low energies.

In Fig.\ \ref{csplot1} we show microscopic and phenomenological elastic scattering cross sections for neutron projectiles on a $^{40}$Ca target at energies of $E=3.2, 5.3, 6.52$\,MeV as well as $^{48}$Ca at  $E=7.97$\,MeV and compare to experimental data \cite{Ca48e7,Ca40e5and6,Ca40e3}. Interestingly, there is very little difference between the predictions from the chiral optical potential and those from phenomenology, where we note that for both results we have added the effects of compound nucleus elastic scattering. The green curves in Fig.\ \ref{csplot1} demonstrate that the Koning-Delaroche global optical potential is in very good agreement with experimental data in this energy regime when both the direct and compound contributions to the elastic scattering cross section are accounted for (cf.\ Ref.\ \cite{Rotureau18}). The compound contribution to the elastic scattering cross section is experimentally indistinguishable to the shape elastic contribution and must be included when comparing to experimental data. In the top plot of Fig.\ \ref{csplot1}, we provide a comparison to the results found in Ref.\ \cite{PhysRevLett.123.092501} for elastic n-$^{40}$Ca scattering at $E=3.2$\,MeV. The results by Idini et al. are obtained through an ab initio calculation of the optical potential using a self consistent Green function approach. We see that the nuclear matter approach in the improved local density approximation gives better agreement with data than the fully ab initio approach of Idini et al., which might be due to different theoretical nuclear density distributions or density of states in the two approaches. In the second plot from the top, we also show that the cross section resulting from the chiral interaction with only two-body contributions is in overall worse agreement with experiment than the chiral potential including both two- and three-body interactions. In the bottom plot we show results of the n-$^{48}$Ca chiral optical potential derived from symmetric nuclear matter. This allows us to better understand the effects of including isospin asymmetry in the chiral optical potential. Overall, we find that the inclusion of isovector terms improves agreement with experiment.

\subsection{Microscopic optical potential at medium-low energy}
In Fig.\ \ref{csplot2} we plot microscopic and phenomenological differential elastic scattering cross sections for neutrons on $^{40,48}$Ca targets at $E = 11.9, 16.9$\,MeV and compare to experimental data \cite{Ca48e12and17,Ca40e12,Ca40e17}. At the neutron projectile energy of $11.9$\,MeV, we find a significant discrepancy between the microscopic results and experimental data at certain scattering angles. In particular, for $E = 11.9$\,MeV the n-$^{40,48}$Ca cross sections from the chiral optical potential have a sharp dip around $\theta=45^\circ$ which is not present in the experimental data. For larger scattering angles, the chiral optical potential results have better agreement with experiment than the KD potential, whose predictions are uncharacteristically departed from experimental data. At $E = 16.9$\,MeV the phenomenological and microscopic optical potentials both predict a dip just below $\theta=40^\circ$ that is partly confirmed by experiment. At larger scattering angles, results from the chiral optical potential tend to overestimate the elastic scattering cross sections, while phenomenological optical potentials moderately underestimate them. The large disagreement between microscopic calculations and experimental results in this narrow energy range may be due to resonances and surface effects that are not accounted for in the nuclear matter approach. One such resonance present in the relevant energy range is the giant dipole resonance (GDR). The cross section for $^{40}$Ca(n,$\gamma$)$^{41}$Ca is shown in Ref.\ \cite{GDR} to be enhanced around $E = 12-20$\,MeV due to the GDR. This resonance could be in part responsible for the large discrepancies between experimental data and our microscopic nuclear matter calculations. In the left two plots we also show that the cross sections resulting from the chiral interaction with only two-body contributions are in worse agreement with experiment than the chiral potential including both two- and three-body interactions. In the right two plots we show results of the n-$^{48}$Ca chiral optical potential derived from symmetric nuclear matter to show the effects of the isovector term of the optical potential. In general, the inclusion of the isovector term improves agreement with experiment.

We also plot the vector analyzing power for $^{40}$Ca at $E = 11.9, 16.9$\,MeV in Fig.\ \ref{ayplot}. The vector analyzing power is a spin observable  defined by
\begin{equation}
A_y(\theta)= \frac{1}{p_y} \frac{\sigma(\theta) - \sigma_0(\theta)}{\sigma_0(\theta)}, 
\end{equation}
where $\sigma$ and $\sigma_0$ correspond to the scattering cross sections for a polarized and unpolarized beam respectively and $p_y$ is the beam polarization in the direction normal to the scattering plane. This quantity is largely determined by the spin-orbit term of the optical potential. In this first direct test of our chiral spin-orbit potential we find that overall it reproduces experimental data well. In particular, the angles at which the polarized cross section $\sigma$ is equal to the unpolarized cross section $\sigma_0$ are reproduced very well.

\subsection{Microscopic optical potential at medium-high energy}
\begin{figure}[t]
\includegraphics[scale=0.35]{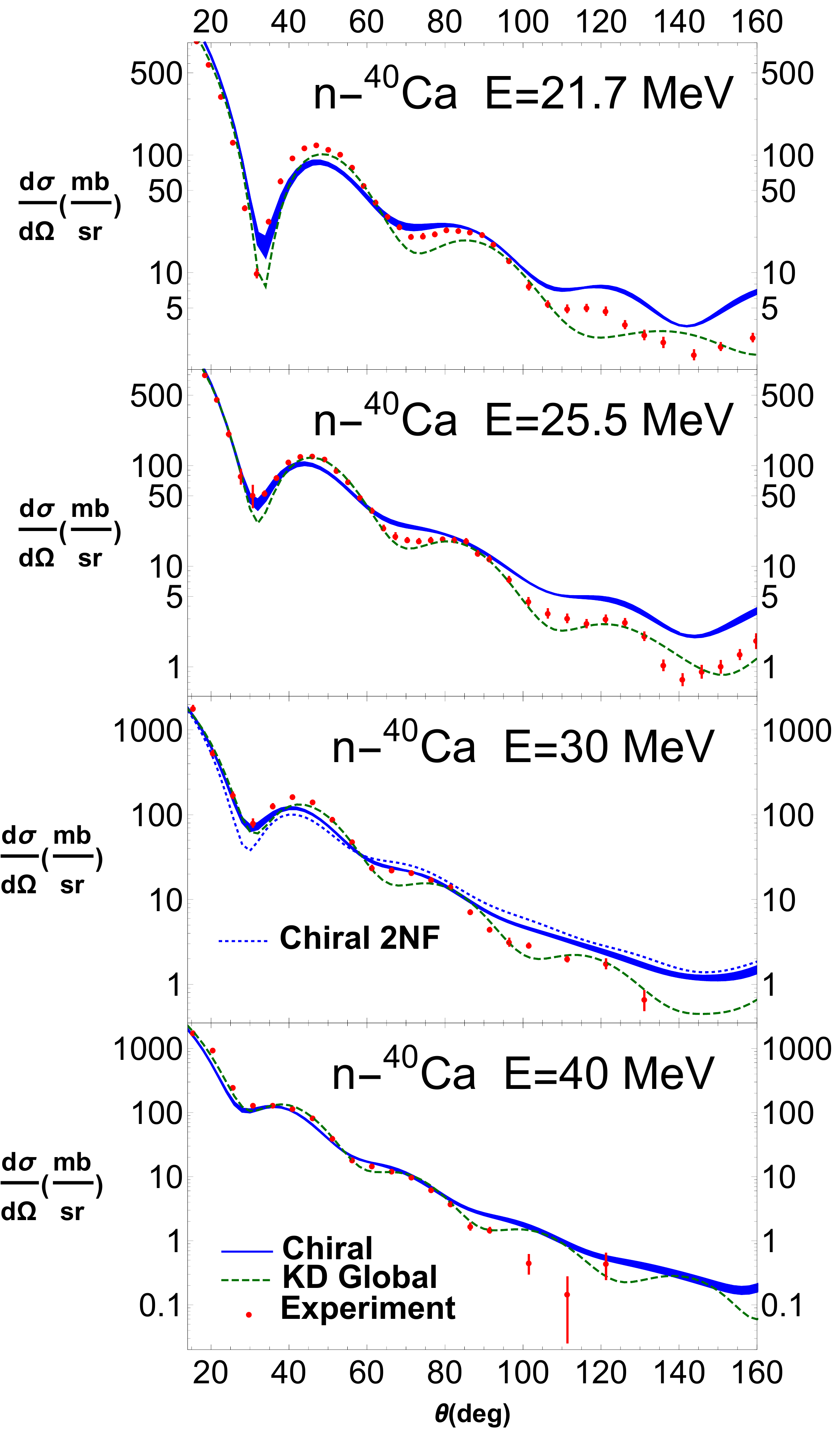}
\caption{Differential elastic scattering cross sections for n-$^{40}$Ca at projectile energies $E=21.7, 25.5, 30, 40$\,MeV. Cross sections calculated from the chiral optical potential are given by blue bands. Cross sections calculated from the Koning-Delaroche phenomenological optical potential are given by green dashed curves, and experimental data are represented by red circles with error bars. In the plot second from bottom, the blue dotted curve is calculated from the chiral optical potential with only two nucleon forces (2NF).}
\label{csplot3}
\end{figure}

In Fig.\ \ref{csplot3} we plot microscopic and phenomenological differential elastic scattering cross sections for neutrons on $^{40}$Ca targets at $E = 21.7, 25.5, 30, 40$\,MeV and compare to experimental data \cite{Ca40e21and25,Ca40e30and40}. For relatively low scattering angles in the range of $0^\circ$ $ < \theta < $ 80$^\circ$, the microscopic optical potential produces cross sections that are consistent with experiment and the phenomenological KD optical potential. For larger scattering angles the microscopic calculations of the cross sections overestimate experimental data and exhibit a weaker interference pattern that persists as the energy increases.

From Fig.\ \ref{fig:potentialplots}, we expect the cause of these discrepancies is the imaginary part of the microscopic optical potential. The microscopic surface imaginary peak is very small in this energy range, as can be seen in Fig.\ \ref{fig:potentialplots}. This leads to larger elastic scattering cross sections. In contrast the imaginary volume part is much larger than phenomenology at higher projectile energies. We have verified that replacing only the microscopic imaginary part with the Koning-Delaroche phenomenological imaginary part leads to significantly improved angular distributions for $\theta > 80^\circ$. In the plot second from the bottom of Fig.\ \ref{csplot3}, we show that the cross section resulting from the chiral interaction with only two-body contributions is in slightly worse agreement with experiment across all angles, but there is not a large difference between the two chiral potentials. This is mainly due to $E=30$\,MeV being close to the energy at which the imaginary terms of the two chiral potentials are approximately equal, as shown in Fig.\ \ref{fig:potentialplots}.

\subsection{Microscopic optical potential at high energy}
\begin{figure}[t] 
\includegraphics[scale=0.37]{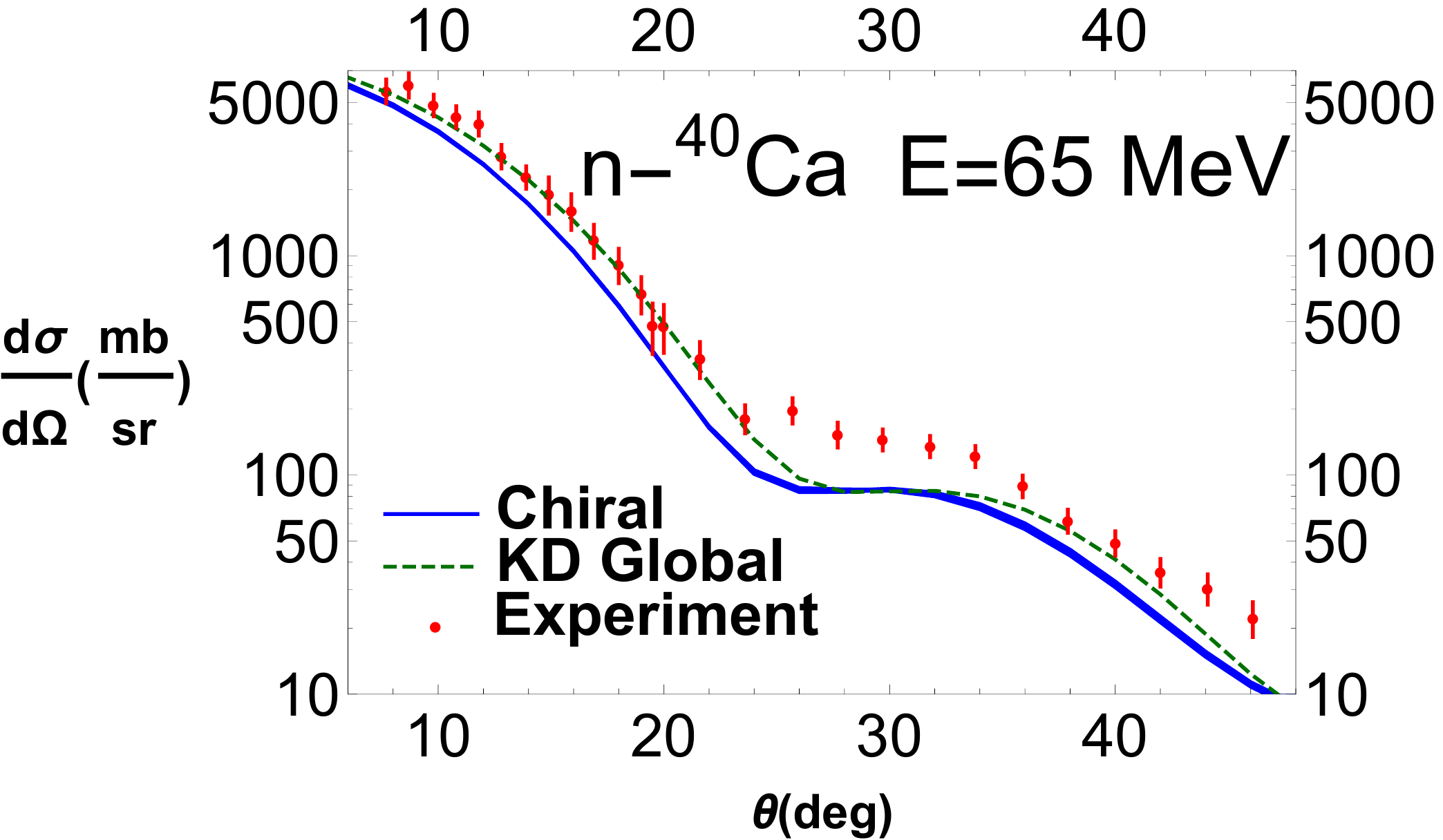}
\caption{Differential elastic scattering cross sections for n-$^{40}$Ca at projectile energies $E=65$\,MeV. The cross section calculated from the chiral optical potential is given by the blue band. The cross section calculated from the Koning-Delaroche phenomenological optical potential is given by the green dashed curve, and experimental data are represented by red circles with error bars.}
\label{csplot4}
\end{figure}

\begin{figure}[t] 
\includegraphics[scale=0.34]{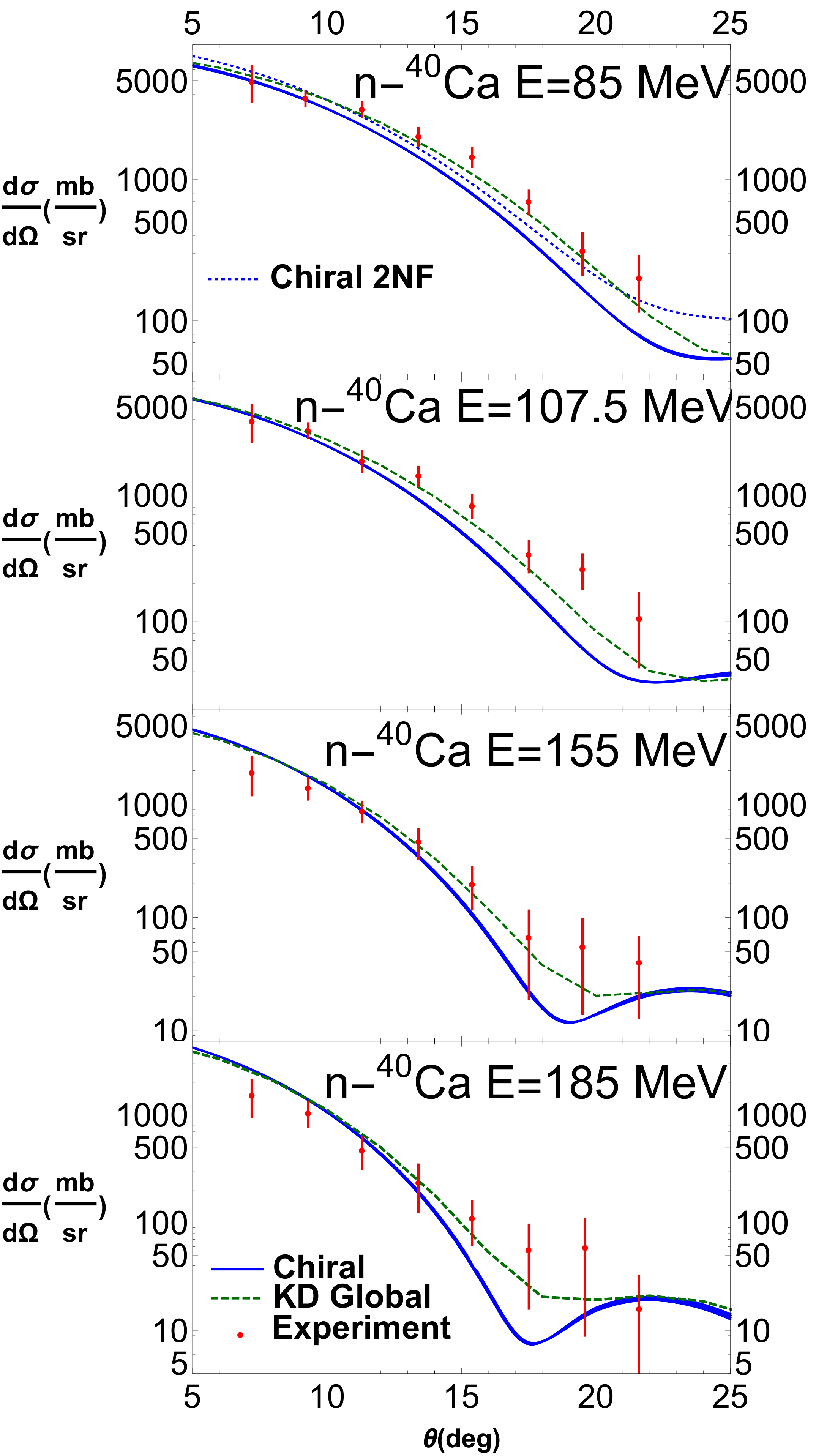}
\caption{Differential elastic scattering cross sections for n-$^{40}$Ca at projectile energies $E=85,107.5,155,185$\,MeV. Cross sections calculated from the chiral optical potential are given by blue bands. Cross sections calculated from the Koning-Delaroche phenomenological optical potential are given by green dashed curves, and experimental data are represented by red circles with error bars. In the top plot, the blue dotted curve is calculated from the chiral optical potential with only two nucleon forces (2NF).}
\label{csplot5}
\end{figure}

In Figs.\ \ref{csplot4},\ref{csplot5} we plot microscopic and phenomenological differential elastic scattering cross sections for neutrons on $^{40}$Ca targets at $E=65, 85, 107.5, 155, 185$\,MeV and compare to experimental data \cite{Ca40e65,Ca40eHigh}. In Fig.\ \ref{csplot4}, we see that the cross section from chiral effective field theory exhibits the same angular dependence as the experimental data, but microscopic many-body theory systematically underestimates the cross section across all scattering angles. In contrast, the KD phenomenological optical potential reproduces the experimental cross section up to $\theta=25^{\circ}$ well. For larger scattering angles, however, the phenomenological cross section is smaller than experiment but very similar to the cross section from chiral effective field theory.

In Fig.\ \ref{csplot5} we compare experimental, phenomenological, and microscopic differential elastic scattering cross sections for n-$^{40}$Ca at $85\,{\rm MeV} < E < 185\,{\rm MeV}$. For these projectile energies, the experimental data span only a small range of scattering angles $\theta \leq 25^\circ$. The data also have large uncertainties which are as large as a factor of $2-5$ in the cross section. The results from chiral effective field theory are consistent with data up to experimental error bars in most cases, but the tendency is again for the microscopic optical potential to underestimate the cross sections. In all cases the KD results are within or very close to experimental data. In the top plot of Fig.\ \ref{csplot5} we show that the cross section resulting from the chiral interaction with only two-body contributions is in slightly better agreement with experiment than the results from the chiral potential with three-body contributions. This is likely due to the fact that the imaginary term of the chiral potential with only two-body contributions happens to be smaller in magnitude and closer to a realistic value at these energies as shown in Fig.\ \ref{fig:potentialplots}.

\subsection{Total cross section}
The total cross section is written as the sum of the elastic scattering and reaction cross section:
\begin{equation}
\sigma_T = \sigma_{\rm el} + \sigma_{\rm re}.
\end{equation}
The reaction cross section in particular is expected to be very sensitive to the strength of the imaginary part of the optical potential. Consequently, we expect chiral optical potentials, with their large imaginary volume parts, to produce a large reaction cross section and hence a large total cross section at high energies. At low and moderate energies, the picture is more complicated as demonstrated in Ref.\ \cite{TheBestPaperEver}. At low energies the microscopic surface imaginary part is small and the imaginary volume part is large compared to phenomenological optical potentials. Depending on the energy, the volume integral of the microscopic imaginary part is therefore either larger or smaller than phenomenology and the reaction cross section behaves analogously.

\begin{figure}[t] 
\includegraphics[scale=0.36]{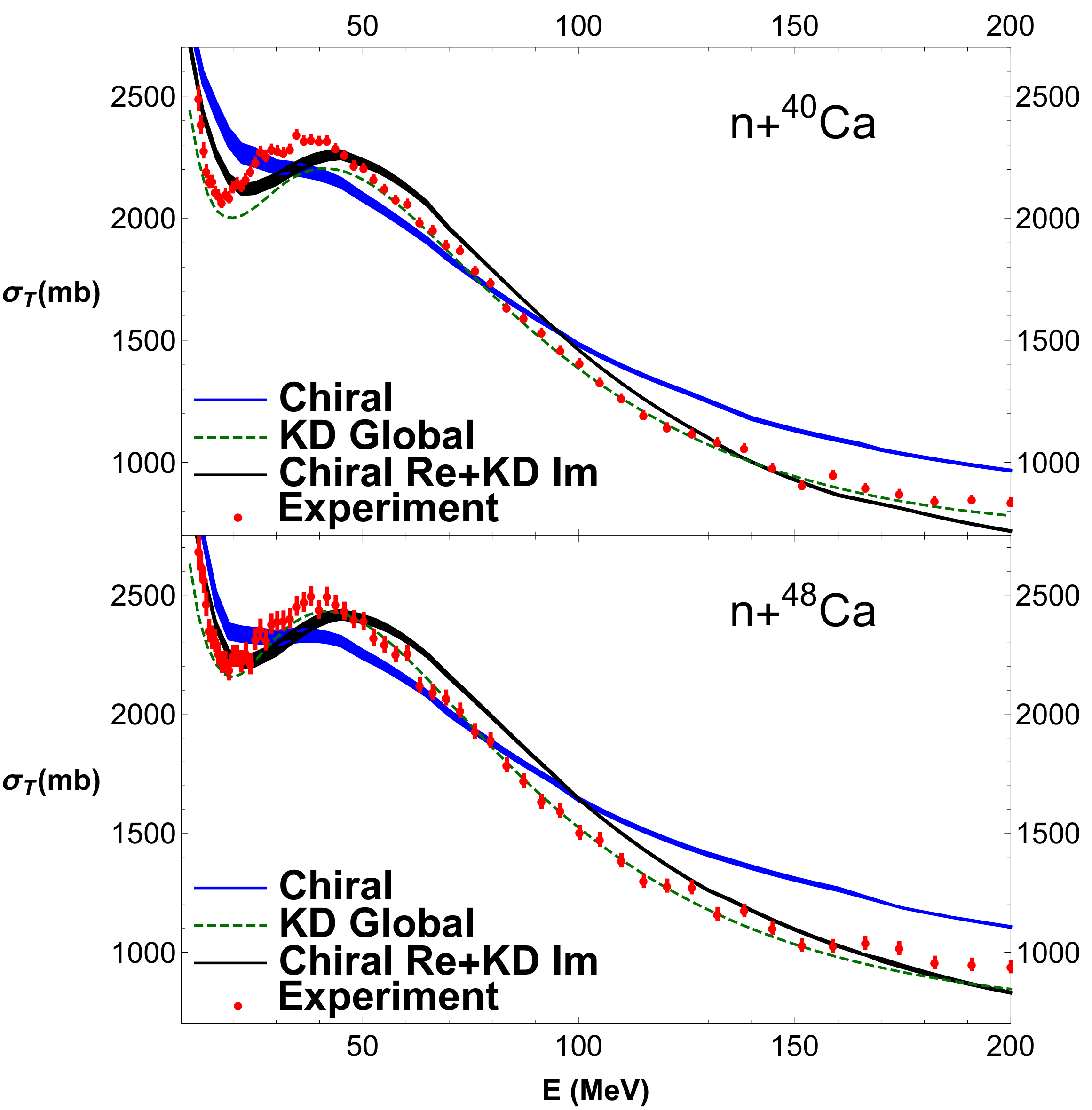}
\caption{The n-$^{40,48}$Ca total cross sections calculated from the chiral optical potential are shown in blue, and the results of the real chiral optical potential plus a phenomenological imaginary term from the Koning-Delaroche optical potential are given by the black curve. Phenomenological results from the Koning-Delaroche optical potential are represented by dashed green curves. Experimental data are shown as red circles with error bars.}
\label{totalcsfig}
\end{figure}

In Fig.\ \ref{totalcsfig} we show the total cross sections for neutron scattering on $^{40,48}$Ca from the chiral optical potential and the KD phenomenological optical potential. In both plots of Fig.\ \ref{totalcsfig} the microscopic optical potential overestimates the total cross section for low energy then underestimates the cross section for medium energy. Past $E=100$\,MeV the total cross section from chiral nuclear optical potentials is systematically too large. As mentioned above, this can be traced to the overly absorptive imaginary term. Overall, the phenomenological optical potential of Koning and Delaroche gives a good description for both isotopes at most energies. The only exception is the n-$^{40}$Ca total cross section for projectile energies in the range $10\,{\rm MeV} < E < 50\,{\rm MeV}$, where the KD total cross sections are small compared to experiment. The experimental data in Ref.\ \cite{TotalCS} were not included in the parameterization of the KD potential since the experiment was performed more recently. Additionally, for the previously mentioned energy range, these experimental data are in slight disagreement with previous experimental results \cite{Finlay93} that the KD potential is fit to. We choose to plot only the more recent data set since total cross section measurements of both $^{40,48}$Ca are made in the same work. In Fig.\ \ref{totalcsfig} the solid black curve is obtained by substituting the KD imaginary part into the chiral microscopic optical potential. We see that indeed there is a significant improvement in the description of the total cross section, which motivates the need to improve the imaginary part of the microscopic optical potential. 

\section{Conclusions}
\label{conclusions}
This work represents the continuation of an effort to construct improved microscopic optical potential based on nuclear two- and three-body interactions from chiral effective field theory. By calculating the nucleon optical potential in nuclear matter for arbitrary density and isospin-asymmetry, one can derive an optical potential for many isotopes across the nuclear chart by utilizing the improved local density approximation. Ultimately our goal is to develop a new generation of microscopic global optical potentials with quantified uncertainties. In previous works the optical potential was calculated in nuclear matter \cite{Holt13omp,Holt15omp} and more recently proton optical potentials were calculated for a chain of calcium isotopes \cite{TheBestPaperEver}. New to this work are calculations of the neutron optical potential for $^{40,48}$Ca and a direct test of the microscopic spin-orbit term by calculating spin observables. We have also compared neutron elastic scattering cross sections from chiral optical potentials with/without three-body forces and with/without isovector terms that demonstrate the importance of including these contributions. 

Overall, we find good agreement with experimental differential elastic scattering data, except in energy regions where unresolved resonances are expected to be important. At the highest energies ($E \simeq80-200$\,MeV) we also find that the large imaginary volume contribution from microscopic optical potentials tends to suppress elastic scattering compared to experimental data. This feature is enhanced in microscopic calculations of the total cross section, which are too large at high energies due to the large reaction cross section induced by the strongly absorptive imaginary part. We have also computed for the first time in our improved local density approximation approach the vector analyzing power. We find that the analyzing power for n-$^{40}$Ca at medium energies is well described by our microscopic optical potentials, validating in particular its spin-orbit part.

We emphasize that no parameters in the model were tuned to experimental reaction data, and therefore the present work together with Ref.\ \cite{TheBestPaperEver} demonstrates the viability of using nucleon-nucleus microscopic optical potentials in regions of the nuclear chart that are unexplored experimentally. In the future we plan to compute neutron-capture cross sections on exotic isotopes and more thoroughly explore theoretical uncertainties \cite{reinert18,Sammarruca18} associated with the isovector part of the nuclear optical potential. We also plan to consider higher-order perturbative contributions to the self-energy that may improve the description of the imaginary part of the optical potential and the overall spin-orbit strength. Lastly, we intend in future works to include several chiral potentials calculated to different orders and with a varying cutoff scales to determine the theoretical uncertainties from the chiral interactions.

\begin{acknowledgments}
We thank F.\ Nunes, G.\ Potel, and J.\ Rotureau for useful discussions. Work supported by the National Science Foundation under Grant No.\ PHY1652199 and by the U.S.\ Department of Energy National Nuclear Security Administration under Grant No.\ DE-NA0003841. Portions of this research were conducted with the advanced computing resources provided by Texas A\&M High Performance Research Computing. Y.L. was supported by the Max Planck Society and the Deutsche Forschungsgemeinschaft (DFG, German Research Foundation) -- Project-ID 279384907 -- SFB 1245.
\end{acknowledgments}

\end{document}